\documentclass[aip,pof]{revtex4-1}
\usepackage{textcomp,amssymb,color,graphicx} 
\usepackage{bm}

\begin{document}

\title{Pairwise hydrodynamic interactions and diffusion in a vesicle suspension} 
\author{Pierre-Yves Gires}
\altaffiliation[Now at ]{Laboratoire de Biom\'ecanique et de Bioing\'enierie, UMR CNRS 7338,
Universit\'e de Technologie de Compi\`egne, France.}
\author{Aparna Srivastav}
\altaffiliation[Now at ]{Zentralinstitut f\"ur Medizintechnik, Technische Universit\"at M\"unchen, Germany}
\thanks{both first authors equally contributed to the production of the data presented in this paper. P.-Y. G. produced the theoretical and computational results and A. S. performed the experiments.}
\author{Chaouqi Misbah}
\author{Thomas Podgorski}
\author{Gwennou Coupier}\email{gwennou.coupier@ujf-grenoble.fr}
\affiliation{Laboratoire Interdisciplinaire de Physique, CNRS - UMR 5588, Universit\'{e} Grenoble I, B.P. 87, 38402 St Martin d'H\`{e}res Cedex, France}

\date{\today}

\begin{abstract}

The hydrodynamic interaction of two deformable vesicles in shear flow induces a net displacement, in most cases an increase of their distance in the transverse direction. The statistical average of these interactions leads to shear-induced diffusion in the suspension, both at the level of individual particles which experience a random walk made of successive interactions, and at the level of suspension where a non-linear down-gradient diffusion takes place, an important ingredient in the structuring of suspension flows. We make an experimental and computational study of the interaction of a pair of lipid vesicles in shear flow by varying physical parameters, and investigate the decay of the \textcolor{black}{net lateral displacement} with the distance \textcolor{black}{between the streamlines on which the vesicles are initially located}. This decay and its dependency upon vesicle properties can be accounted for by a simple model based on the well established law for the lateral drift of a vesicle in the vicinity of a wall. In the semi-dilute regime, a determination of self-diffusion coefficients is presented.

\end{abstract}


\maketitle

\section{Introduction}

Liquid suspensions of deformable particles are the focus of permanent interest due to their ubiquity in life science and applications, from emulsions to blood, a dense suspension of red blood cells. It is well known since Batchelor \cite{batchelor72c,zinchenko84} that the viscosity of a semi-dilute suspension of rigid spheres departs from the classic linear Einstein law of viscosity for volume fractions of particles above a few percents, due to the additional dissipation induced by hydrodynamic interactions between particles.

In addition to their influence on the effective viscosity at significant volume fractions, these \textcolor{black}{hydrodynamic interactions  (which are sometimes called binary collisions in the literature, although still mediated by hydrodynamics)} can lead to irreversible perturbations of the particle trajectories which result in an effective random walk of individual particles in the suspension. This shear-induced diffusion has two main consequences: enhanced mixing and transport even at low Reynolds number \cite{goldsmith71,goldsmith79,cha01,higgins09,zhao11,tan12}, and a modification of the structure of suspensions via diffusion along gradients of concentration of the particles \cite{hudson03,rusconi08,podgorski11,grandchamp13}.

In shear flow, two identical particles located on different streamlines and moving towards each other will generally experience irreversible drift in the shear and vorticity directions after they have interacted. However, for smooth rigid spherical particles in a dilute regime dominated by pairwise interactions, the cross-stream lateral displacement is expected to be negligible at low Reynolds number since trajectories must be symmetric due to the flow-reversal symmetry of the Stokes equation \cite{bretherton62} and the symmetry of the geometrical configuration. Symmetry breaking can be obtained by considering rough particles \cite{dacunha96,blanc11} or deformable two-fluid systems such as bubbles \cite{wijngaarden76} or drops \cite{loewenberg97,guido98,singh09,le11}. More recently, systems made of closed membranes have been investigated numerically or experimentally. Hydrodynamic interaction between elastic capsules were studied numerically in several papers  \cite{lac07, lac08,pranay10,omori13}. During the interaction, net displacement of the capsules can be coupled with wrinkling or buckling of the membranes, whose tension strongly increases during interaction.

\begin{table*}

  \begin{ruledtabular}

    \begin{tabular}{clcc}

	Set&Solution  & $\eta$ (mPa.s) & $\lambda=\frac{\eta_{I}}{\eta_{E}}$ \\ \hline

   1 & \begin{tabular}{l}(I) 300mM sucrose in (20$\%$ glycerol + 80$\%$ water w/w) \\

    (E) 370mM glucose in (20\% glycerol + 80$\%$ water w/w) \end{tabular} & \begin{tabular}{c} 2.2 \\ 2.2 \end{tabular} &1.0 \\ \hline

    2 & \begin{tabular}{l}(I)  100mM sucrose in water + 3.3$\%$ dextran w/w \\

   (E) 115mM glucose in water \end{tabular} & \begin{tabular}{c} 4.2 \\ 1.1 \end{tabular} & 3.8 \\  \hline

     3& \begin{tabular}{l}(I)  300mM sucrose  in water\\

   (E) 316mM glucose in water + 3$\%$ dextran w/w \end{tabular} & \begin{tabular}{c} 1.1 \\ 4.0 \end{tabular} & 0.28 \\     \end{tabular}

    \end{ruledtabular}

  \caption{Sets of internal (I) and external (E) solutions  considered in the experiments. Viscosities $\eta$ are measured at  $T = 23^\circ$ C.}

  \label{tab:sol}

\end{table*}

The dynamics and rheology of suspensions of lipid vesicles have recently been the focus of several studies, due to their relevance to the understanding of blood flows, considering giant vesicles as models of red blood cells, and the challenging theoretical questions they pose as a consequence of their rich microscopic dynamics. Vesicles are closed lipid bilayers with mechanical properties similar to those of living cells. A key property is the membrane inextensibility, which leads to  local area conservation, while volume conservation is generally obtained once osmotic equilibrium is reached. The vesicles mechanical response, as well as their shapes \cite{helfrich73},  are governed by a bending energy of order a few $kT$, {where $k$ is Boltzmann's constant and $T$ the temperature}. These particular properties, especially the non-linearities due to the constraint of local area conservation, are responsible for the \textcolor{black}{various} dynamics of single vesicles in shear flow \cite{deschamps09b,farutin10,Biben11}. The phase diagram of microscopic dynamics has a signature on the rheology of vesicle suspensions \cite{danker07a}, but there is still disagreement, especially in the semi-dilute regime where two experimental studies show contradictory results \cite{kantsler08,vitkova08}. In an effort to resolve this contradiction, Kantsler et al. \cite{kantsler08} and Levant et al. \cite{levant12} have investigated the influence of interactions on fluctuations and correlations of the inclination angles of interacting vesicles and suggest that they may be responsible for discrepancies between theories in the dilute regime and experimental measurements of the effective viscosity, which are often made in a semi-dilute regime for sensitivity reasons. On the analytical side, the trajectories of interacting vesicles  have been recently studied in the limit were they are initially very distant from each other \cite{gires12}. This study has been later on refined in the case of vesicles located in the same shear plane \cite{farutin13}. Very recently, such  trajectories have been calculated numerically  by Zhao and Shaqfeh \cite{zhao13}. They also calculated the rheology of a semi-dilute suspension and found good agreement with the experiments by Vitkova et al. \cite{vitkova08}, a strong indication that, in that concentration regime, interactions between vesicles cannot explain the contradiction between the latter experiments and the one by Kantsler et al. \cite{kantsler08}

Along with their influence on rheology, hydrodynamic interactions significantly affect the structure of suspensions, especially in confined flows where a balance between migration away from walls and shear-induced diffusion due to repulsive interactions leads to the formation of a non-homogeneous distribution of vesicles \cite{podgorski11}. During heterogeneous interactions of vesicles or capsules with different mechanical or geometrical characteristics, asymmetric displacements take place, which leads to segregation or margination \cite{podgorski11,zhao11,zhao12,kumar12,fedosov12}, a phenomenon also observed in blood flows \cite{goldsmith84,tilles87,Uijttewaal93,yeh94}.

In this paper, we report on our experimental and numerical investigation of the interaction of two identical vesicles in shear flow, with a focus on the net lateral displacement as a function of initial configuration and vesicle properties.

With a good agreement between experiments and simulations, the amplitude of the lateral displacement is found to be weakly dependent on vesicle deflation and viscosity ratio, at least in the tank-treading regime to which we restrict our study. Thanks to the simulations, we also discuss to which extent the discrepancies between the ideal case of two identical and neutrally buoyant vesicles, placed in the shear plane of an infinite simple shear flow, and the realistic case of channel flow, influence the final result.

Finally, from the numerical results, an evaluation of the self-diffusion coefficient, obtained by averaging displacements over all initial configurations, is proposed and compared to the recent results of Zhao and Shaqfeh \cite{zhao13}.

\begin{figure}[t!]

\includegraphics[width=\columnwidth]{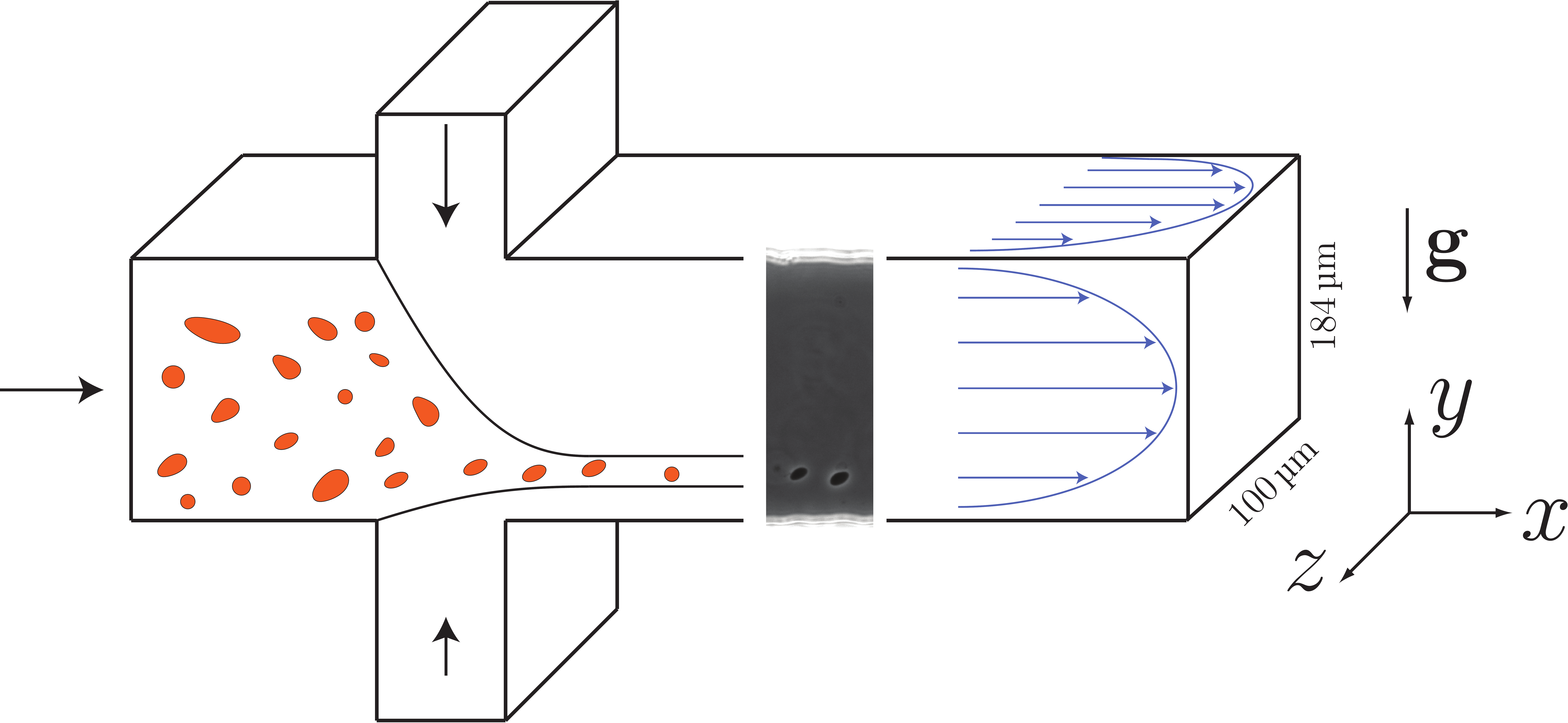}

\caption{Sketch of the experiment.}\label{sketch_exp}

\end{figure}

\section{Experimental set-up}

Fluid vesicles are prepared by following the electroformation method \cite{angelova92}, which produces vesicles of various size and deflation {(that is, the surface to volume ratio)}. They are made of a dioleoylphosphatidylcholine (DOPC) lipid bilayer. We consider three sets of outer and inner solutions in order to vary the viscosity ratio $\lambda$ between the inner and the outer fluids (see table \ref{tab:sol}). The different additives (sugars and dextran) used for inner and outer solutions provide an optical index contrast which is convenient for phase contrast microscopy.

We wish to observe interactions in simple shear flow between vesicles located in the same $xy$ plane, where $x$ is the flow direction  and $y$ the shear direction. To that end, the vesicle suspension is injected in a standard polydimethylsiloxane (PDMS) microfluidic device. The observation channel is 184\,{\textmu}m wide ($y$ direction) and 100\,\textmu m deep ($z$ direction). The imposed flow is along the $x$ axis (see Fig. \ref{sketch_exp}).  Before the observation section, vesicles flow in a channel of several centimeters long, so that centering in the $z$ direction is generally rather well achieved \cite{coupier08}, as  confirmed by the location of all vesicles within a focal plane of thickness of order 5\,\textmu m. The interacting vesicles were followed manually translating the stage. The observation window is 477x358 $\text{\,\textmu m}^2$, with a resolution of $0.47$\,\textmu m/pixel. The use of a channel flow, rather than a four-roll mill device\cite{kantsler08,levant12}, allows to measure the final lateral displacement due to the interaction, a key parameter in the discussion of diffusion phenomena.

As  measurable interactions only occur when vesicles  have initial $y$ separation not larger than 2 radii, it appeared necessary to favor such an initial condition by adding a flow focusing device at the entrance of the observation channel. Two lateral inlets were then added, where vesicle-free fluid was injected in order to focus the suspension in a narrow area. This area is located at around one fourth of the total width of the channel, that is, far from the wall and far from the center, where the flow can be considered as a simple shear flow, in a first approximation {to be discussed later}. Vesicles stay at this {favorable} position thanks to the balance between lift forces \cite{abkarian02,callens08,coupier08} and gravity. In addition, dilution by the lateral inlets decreases the probability of perturbation of the interaction trajectories by other vesicles.

\begin{figure*}[t!]

  \includegraphics[width=\columnwidth]{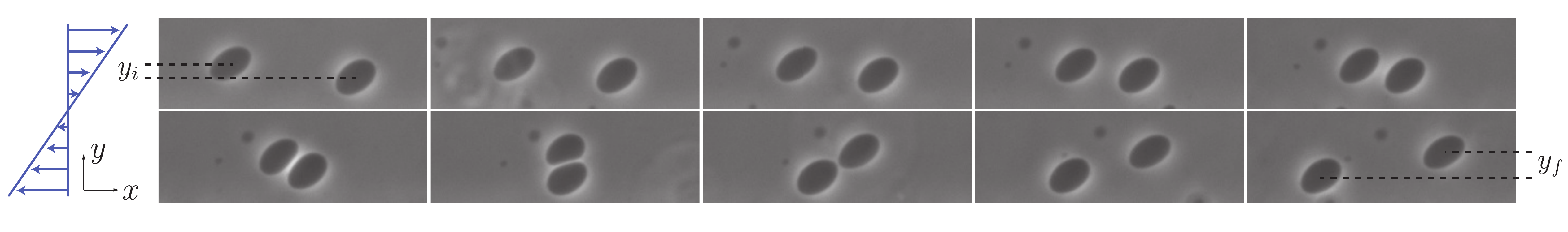}

\caption{Time sequence of an experimental interaction (left to right and top to bottom). Bottom vesicle: $R_1=9.3\,$\textmu m, $\nu_{a1}=0.94$. Top vesicle: $R_2=9.2\,$\textmu m, $\nu_{a2}=0.92$. Total sequence length is about 5\,s.\label{fig:trajexp}}

\end{figure*}

In the observation window, at most 3 or 4 vesicles (including the two studied vesicles) are present at the same time. In the selected interaction sequences, the additional vesicles of non negligible size are always at a distance from the pair larger than 5 radii \textcolor{black}{ and are located almost on the same streamlines, so that they will not come close to the pair within the duration of the studied interaction process.} As in  Fig. \ref{fig:trajexp}, very small vesicles may come closer, but the induced perturbation is expected to be negligible: from an asymptotic approach\cite{gires12}, we can expect the velocity perturbations induced by a vesicle 4 times smaller than the studied ones to be smaller than the one coming from the interacting vesicles by a factor $({\frac{1}{4}})^2=0.06$.

Once  an appropriate pair of vesicles is chosen, the vesicles are followed along their trajectories and the $(x,y)$ coordinates of the vector linking their geometrical centers are determined, as well as their shapes. We denote by $(x_i,y_i)$ the initial position and by $(x_f,y_f)$ the final one. By convention, $x_i<0$ and $y_i>0$. An example of selected snapshots taken along a trajectory is shown in Fig. \ref{fig:trajexp}.  As we only have access  to their two-dimensional cross-section in the $xy$ plane, we characterize the 2D shapes by the effective radius $R_{\text{i}}$, $\text{i}=1,2$, defined by $R_{\text{i}}=\mathcal{P}_{\text{i}}/(2 \pi)$, where $\mathcal{P}_{\text{i}}$ is the cross-section perimeter, and by a reduced area $\nu_{a\text{i}}=\mathcal{A_{\text{i}}}/(\pi R_{\text{i}}^2) \le 1$, where $\mathcal{A_{\text{i}}}$ is the cross-sectional area. These two parameters are evaluated before the vesicles strongly interact, at which point out-of-plane deformations occur.

In this study, we focus on pairs of vesicles of similar size and deflation (within maximal variations of 10 percent for the radii and 5 percent for the reduced area). We denote by $R_0$ and $\nu_a$ the arithmetic averages of the radii and reduced areas of the two interacting vesicles. $R_0$ lies between 5 and 19\,\textmu m, and  $\nu_{a}$ between 0.73 and  1. The flow velocity is set so that the capillary number lies typically between 10 and 100. This capillary number $Ca=\eta\dot{\gamma} a^{3}/{\kappa}$  qualitatively represents the ratio between the magnitude of the liquid viscous stresses exerted on a membrane, and its resisting bending stresses, controlled by  bending rigidity $\kappa$. $\dot{\gamma} $ is the shear rate and $\eta$ the suspending fluid viscosity. $a$ is the effective radius of the vesicle, defined from its volume $V$ by $V=4\pi a^3/3$. Note that, due to vesicle volume conservation, this 3D effective radius is constant and characteristic of the considered vesicle, while the observed 2D radius $R_0$ depends on the applied flow. As $a$ can only be roughly estimated in the experiments, we only have access to estimated values for $Ca$.

From the obtained trajectories, we extract the main information, that is the lateral displacement $\Delta_y=y_f-y_i$ as a function of initial lateral separation $y_i$. Both distances are rescaled by $R_0$. Several initial positions $y_i$ are scanned either by considering different pairs of vesicles, or by making a given pair going back and forth thanks to flow reversal.

\section{Experimental results}

\label{experimental_results}

\begin{figure}[t!]

\includegraphics[width=\columnwidth]{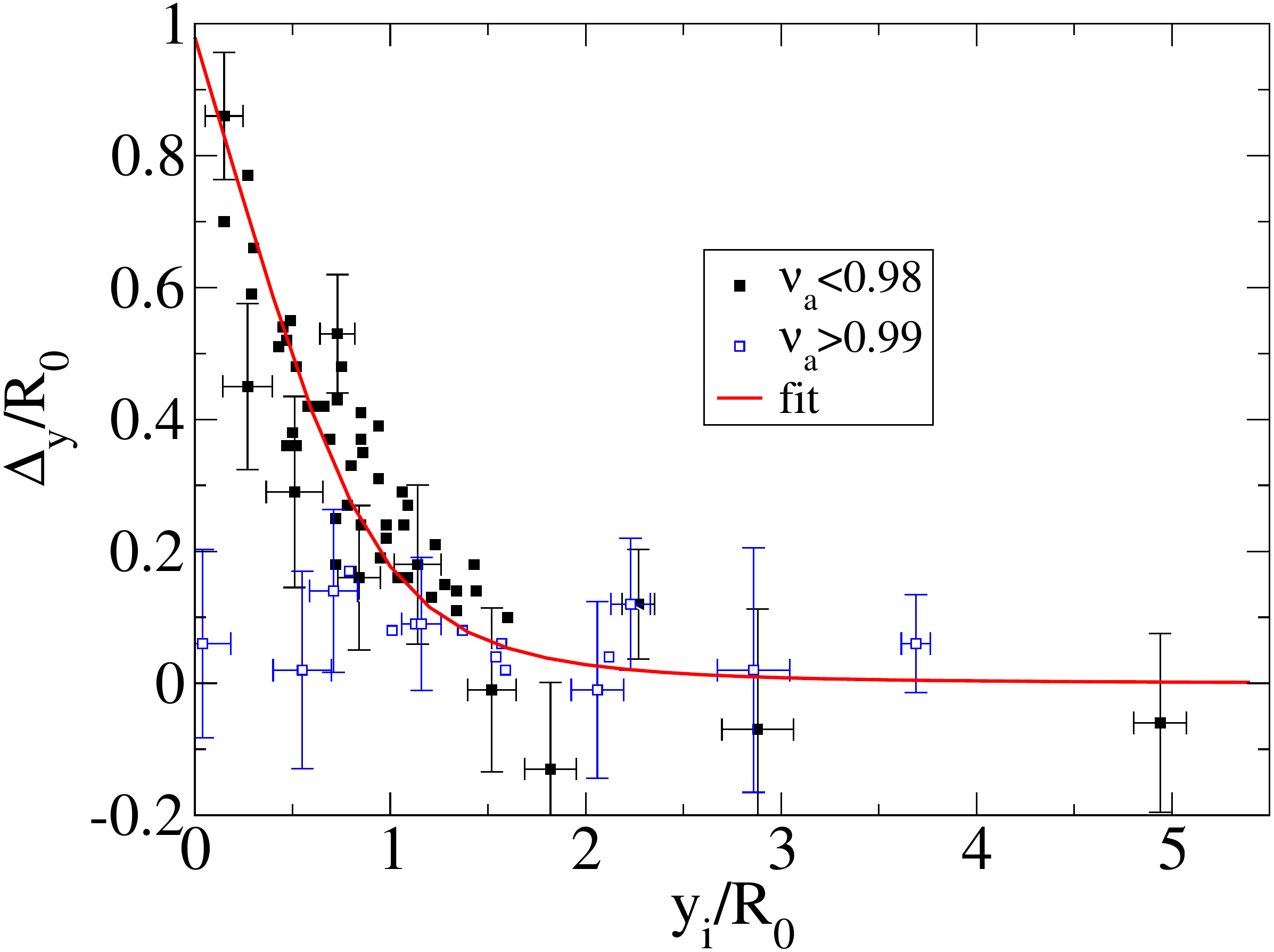}

\caption{Experimental lateral displacements $\Delta_y/ {R}_0$ as a function of $y_i/{R}_0$ for vesicles with no viscosity contrast ($\lambda=1$). Empty symbols correspond to vesicles with reduced area $\nu_a > 0.99$, while vesicles of reduced area between 0.82 and 0.98 are represented by full symbols. Full line shows fit to this latter data set with empirical law given by Eq. \ref{eq:emplaw} with fit parameter $\xi=0.92$. {Error bars correspond to a spatial uncertainty of $1$\,\textmu m. Only part of them are presented for clarity.}\label{fig:exp-lambda1}}

\end{figure}

We first focus on vesicles with no viscosity contrast. Results for $\Delta_y/ {R}_0$ as a function of $y_i/{R}_0$ are shown in Fig. \ref{fig:exp-lambda1}.  Initial and final $y$ positions are measured by averaging over several positions long before and after interaction. Error bars are associated to the fluctuations in these $y$ positions due to the presence of other small vesicles or flow variations due to channel roughness. Such events are likely to occur because of the large ratio between the relative velocity along the $x$ axis between the considered vesicles and the other vesicles or the wall, and the velocity along the $y$ axis. The studied pairs are split into two subpopulations according to their reduced areas.  Vesicles with $\nu_a > 0.99$ undergo negligible deviations which are not measurable within experimental errors. This result is expected for spherical particles and allows to check that no uncontrolled drift \textcolor{black}{alters} the experimental results. All other pairs of  vesicles yield comparable deviations whatever the reduced area  in the range  $0.82-0.98$. Data scattering can be due to variations in reduced areas (including within a pair), sizes, capillary numbers, but also to non complete colocation in the same $xy$ plane.  In addition, displacements might be affected by the flow perturbation induced by the walls, which depends on the lateral position of the vesicles, a parameter that varies from one pair to another.

Lateral deviation is a decreasing function of initial lateral separation, and becomes negligible for initial \textcolor{black}{lateral} separations greater than one diameter. An empirical estimate for this deviation can be obtained by considering that, in the reference frame of the bottom vesicle, the displacement of the top vesicle is due to the interaction with a wall of finite extent in the $x$ direction, whose role is played by the bottom vesicle. The lift velocity of a vesicle in a simple shear flow and at a distance $y$ from a wall was experimentally shown \cite{callens08} to agree with the scaling law $\dot{y}=U \dot{\gamma} R_0^3/y^2$ suggested or confirmed by several theoretical works \cite{olla97,sukumaran01,vlahovska07,farutin13}, where $U$ is a dimensionless parameter that depends on viscosity ratio and reduced volume. The reduced volume $\nu$ is defined as the ratio between the vesicle volume and the volume of the sphere having the same area; due to volume conservation and membrane incompressibility, this is a constant parameter that characterizes the deflation of the vesicle. The top vesicle flows with relative velocity  $\dot{x}=\dot{\gamma} y$, so that  $dy/dx=U R_0^3/y^3$. Interaction takes place on a finite distance of order $2R_0$. Integrating the latter equation on this distance for $x$ and between $y_i$ and $y_f$ for $y$, one finally finds

\begin{equation}
\Delta_y/R_0=\big(y_{i}^4/R_0^4+\xi\big)^{1/4} - y_{i}/R_0,\label{eq:emplaw}
\end{equation}

where $\xi= 8 U$ contains the interaction details. $\xi^{1/4}$ is the maximal displacement, obtained for $y_i \to 0$. Taking an estimate of  $U$ from Olla's work \cite{olla97}, we have for a prolate ellipsoid with $\nu_a= 0.9$, $\xi^{1/4}\sim 1.2$, which is close to the maximal displacement seen in Fig. \ref{fig:exp-lambda1}, where a full fit of the whole data set with Eq. \ref{eq:emplaw} yields $\xi= 0.92$. Note that this is a single parameter fit, so that the distance at which interaction becomes negligible is fully determined by the maximal deviation obtained for quasi-aligned vesicles. In particular, vesicles initially distant by one diameter  \textcolor{black}{ in the $y$ direction} will deviate only by $(16+\xi)^{1/4}-2 \lesssim 2\%$  from their initial trajectories.

From this law, we can estimate how the final displacement should vary with reduced volume. {Following for instance the recent study by Farutin and Misbah \cite{farutin13}, $U$ scales as $(1-\nu)^{\frac{1}{2}}$, so that the maximal displacement scales as $(1-\nu)^{\frac{1}{8}}$.} {This sharp increase around $\nu=1$ explains the strong difference between the quasi-spherical vesicles ($\nu_a>0.99$) and the more deflated ones ($\nu_a<0.98$) seen on Fig. \ref{fig:exp-lambda1}. On the other hand, from Olla's results \cite{olla97},  $U$ is multiplied by a factor 2.7 between prolate vesicles of reduced areas 0.98 and 0.82, respectively. The maximum displacement for vanishing $y_i$ should then be multiplied by $2.7^{1/4}\simeq 1.3$. Such a tiny variation is within the scattering and error of experimental data.}

\begin{figure}[t!]

 \includegraphics[width=\columnwidth]{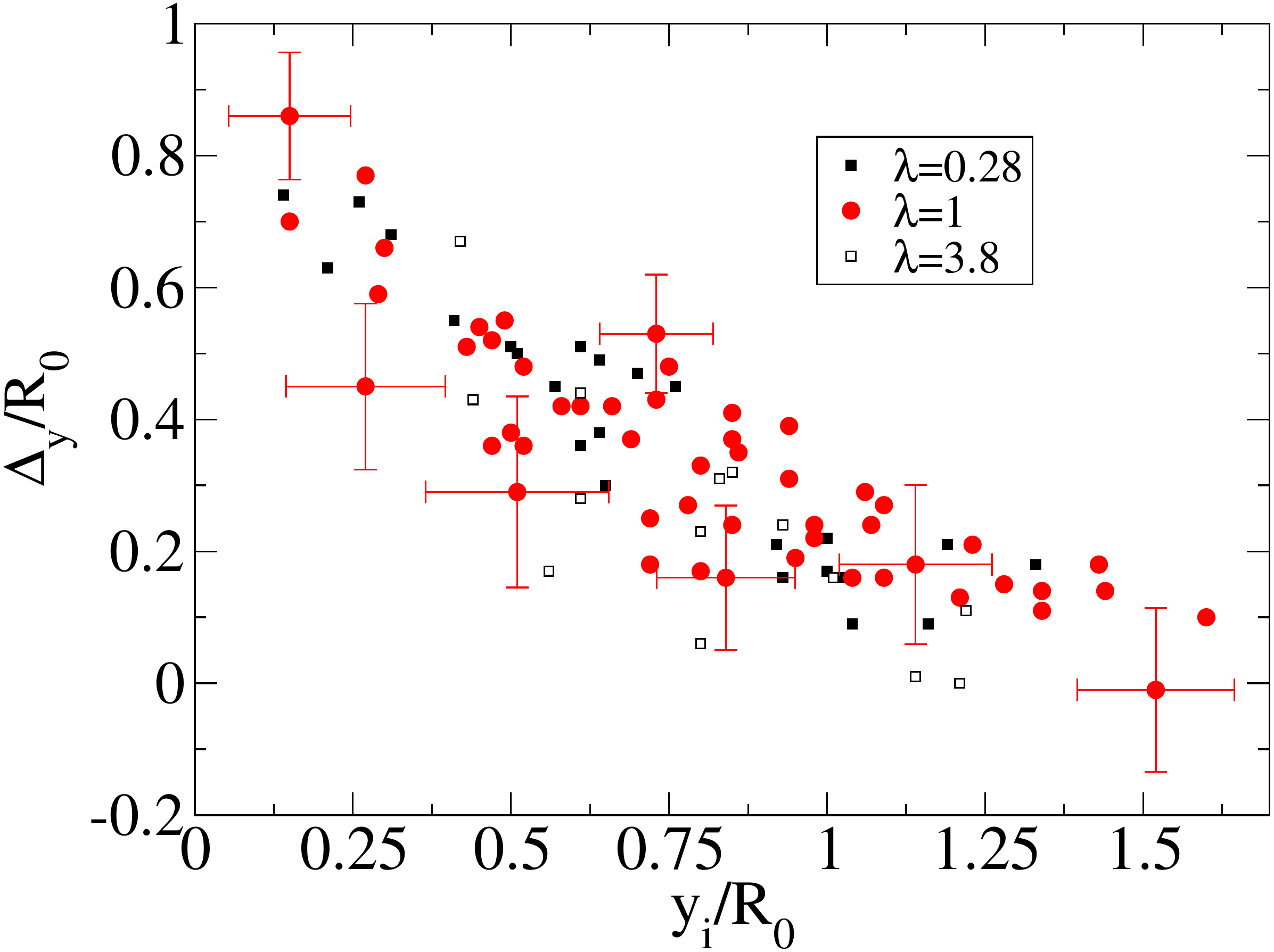}

\caption{Experimental lateral displacements $\Delta_y/ {R}_0$ as a function of $y_i/{R}_0$ for vesicles with different viscosity ratios; $\lambda=0.28$: $\nu_a\in[0.73;0.98]$; $\lambda=1$: $\nu_a\in[0.82;0.98]$ ; $\lambda=3.8$: $\nu_a\in[0.77;0.98] $.  {Error bars correspond to a spatial uncertainty of $1$\,\textmu m. Only part of them are presented for clarity.}\label{fig:exp-lambda-autre}}

\end{figure}

Similarly, when the viscosity ratio is varied, no significant displacement variation is observed, as shown in  Fig. \ref{fig:exp-lambda-autre}. Once again, this is consistent with our empirical law, since from Olla's results again, for $\nu_a=0.9$, $U^{1/4}$ decreases only by  2\%  between vesicles of viscosity ratio  0.28 and 1, and by 12\% between vesicles of viscosity ratio  1 and 3.8.  {According to Zhao and Shaqfeh\cite{zhao13}, the maximum displacement drops  by about 30\% between viscosity ratio 1 and 7.}

In the next section, we address the same questions with full 3D numerical simulations {restricted to the case $\lambda=1$, following our discussion on the weak influence of the viscosity ratio in the previous section. We then confront the numerical results with the experimental ones.}

\section{Model and numerical method}

\subsection{Liquid and membrane}

In this section we outline the model and numerical method.
The internal and external liquids are modeled as incompressible, homogeneous, Newtonian fluids. We restrict our study to the case where their densities, as well as their viscosities, are equal. Both liquids flow in the creeping regime.

The membranes are modeled by two dimensional surfaces. As for the liquids, their inertia is negligible. Their areas stay locally constant. They resist bending with an energy $E_b$, given by \cite{Helfrich}

\begin{equation}
\label{hel}
 E_b=\int_{A}\frac{\kappa}{2}(2H)^{2}dA,
\end{equation}

where $A$ is the membrane surface, $\kappa$ the bending rigidity, and $H$ the mean curvature. The sign convention for the curvatures is taken so that the mean curvature of a sphere is negative.

The resulting surface force density that the membrane exerts on the fluids is

\begin{eqnarray*}
\mathbf{f}=-\{\kappa[2H(2H^{2}-2K)+2\Delta_{s}H]-2\zeta H\}\mathbf{n}+\boldsymbol{\nabla}_{s}{\zeta},
\end{eqnarray*}

where $\mathbf{n}$ is the unit normal vector pointing outward, $K$ the Gaussian curvature, and $\zeta$ a Lagrange multiplier that enters the total energy, obtained by adding to  (\ref{hel}) $\int_{A}\zeta dA$ . It ensures local membrane incompressibility and satisfies:

\begin{equation}
\label{2d_inc}
\boldsymbol{\nabla}_{s}\cdot \boldsymbol{v}=0,
\end{equation}

where $\boldsymbol{\nabla}_{s}$ is the surface gradient operator and $\boldsymbol{v}$ is the membrane velocity.

\textcolor{black}{Note that we don't include in our model any other small range interaction than the hydrodynamic forces within the lubricating film, described as squeezed between athermal membranes. As a first approximation, we considered that theses stresses grow fast enough so that the minimal distance between the membranes, that we denote $d$, remains higher than a typical distance under which other type of interactions become significant. The first one that would appear is linked to the inhibition of thermal fluctuations\cite{dehaas97_2}, which leads to an entropic repulsion pressure. It is of order $0.2(k_{\text{B}}T)^2/(\kappa d^3)$. It would balance the imposed pressure, estimated as $\eta \dot \gamma$, that tends to push the two vesicles towards each other, if
\begin{equation} 
d\sim ({0.2(k_{\text{B}}T)^2\over \kappa \eta \dot\gamma})^{1/3}.
\end{equation}
Using the typical value $\kappa\sim 20 k_{\text{B}}T$, for the smallest shear rate in our experiments $1s^{-1}$, one finds that $d$ reaches values in the range of $100$nm. We checked that, in the trajectories we investigated, $d$ remains higher than the previous estimate. The facts that the entropic force is repulsive, and that, on the contrary to some rigid particles, there are no heterogeneities on the phospholipid membranes that can facilitate the drainage of the lubricating film, support even more our approximation.}

\subsection{Boundary conditions}

The membranes are supposed to be at osmotic equilibrium and are modeled as impermeable. Together with the no slip boundary condition, this leads to an advection of the membranes with the local velocity of the flow.

A force balance on the membrane yields

\begin{equation}
\mathbf{f}=-(\sigma^{+}-\sigma^{-})\cdot \mathbf{n},
\end{equation}
where $\sigma$ is the liquid stress tensor with a $+$ or $-$ superscript respectively for the external and internal fluids, defined as $\sigma=-p1\!\!1+\eta\left(\boldsymbol{\nabla}\boldsymbol{v}+(\boldsymbol{\nabla}\boldsymbol{v})^t\right)$. Far from the vesicles, the imposed simple shear flow  $\mathbf{v}^{\infty}=\dot{\gamma}y\boldsymbol{e}_{x}$ is recovered.

\begin{figure}[h!]

\centering

\includegraphics[width=0.9\linewidth]{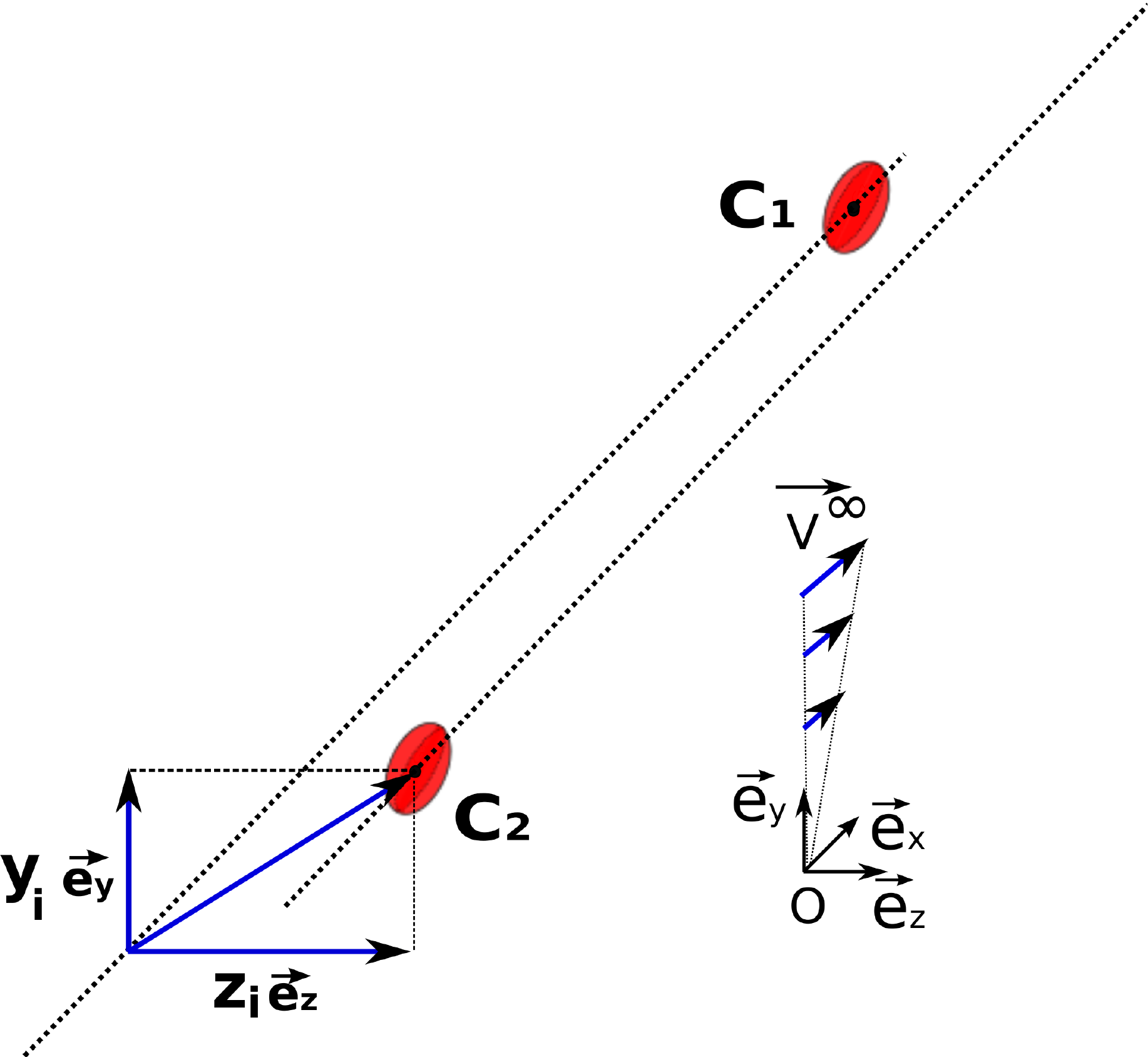}

\caption{\label{sketchsystem}A schematic view of the initial state of the system}

\end{figure}

We denote by $\mathbf{R_{12}}=(x,y,z)$ the vector linking the centers of mass $C_{\text{i}}$ of the two vesicles.  We shall study  the evolution of $(y,z)$ as a function of $x$, that is, the trajectory of vesicle 2 in the frame centered on vesicle 1. Different initial positions $(y_i,z_i)$ will be scanned, with initial longitudinal distance $x_i$ much larger than the vesicles radii. A sketch of the initial state of the system is presented in Fig. \ref{sketchsystem}.

\subsection{Numerical method}

The full set of equations in the Stokes regime can be converted into a boundary integral formulation \cite{Pozrikidis92}. The integral equation (recalled below) is solved numerically in three dimensions following  the work by Biben \textit{et al.}\cite{Biben11}. The new elements of the present study are the extensions to two vesicles and, in a second time, to the presence of a wall that turns out to be a relevant ingredient when confronting the numerical results with the experimental one. We shall first study the situation without wall, which is the main goal of the paper.

The integral equation provides the expression of the membrane velocities as a function of boundary integrals and reads
\begin{eqnarray}
\label{bim_memb}
v_{\alpha}(\mathbf{r})&=&v_{\alpha}^{\infty}(\mathbf{r})+\int_{\partial\Omega}G_{\alpha\beta}(\mathbf{r},\mathbf{r}')f_{\beta}(\mathbf{r}')dA',
\end{eqnarray}

where $\mathbf{r}$ is the position vector of a membrane point, $\partial\Omega$ the boundaries present in the system under consideration, which are in the present case the two vesicle membranes, and $G(\mathbf{r},\mathbf{r}')$ the Green's function of an incompressible fluid following Stokes equation. As we consider an  unbounded domain, an appropriate choice is the Green's function associated to a point force in an infinite liquid, such that $G_{\alpha\beta}(\mathbf{r},\mathbf{r}')=G_{\alpha\beta}^{\infty}(\mathbf{r}-\mathbf{r}')$, where \cite{Pozrikidis92}

\begin{eqnarray}
 G^{\infty}_{\alpha\beta}(\mathbf{r})=\frac{1}{8\pi\eta}\left( \frac{\delta_{\alpha\beta}}{r}+\frac{r_{\alpha}r_{\beta}}{r^{3}} \right).
\end{eqnarray}

For most simulations, the vesicles are meshed by $642$ vertices, and the time step is $10^{-4}{\eta a^3}/{\kappa}$. {We checked that for a typical trajectory ($(y_i,z_i)=(0.5,0)$), results were relatively independent from a reduction of the mesh size and time step: increasing the number of vertices to $2562$ and reducing the time step by a factor 2 led to relative changes in the transverse migration of $0.3\%$.} A challenge is to achieve an evolution of the membrane shapes ensuring a local conservation of the area. We present in appendix \ref{appendixA} details showing that our study conserves
the area with  a good approximation. The simulations start with both vesicles having the steady inclination angle of an isolated vesicle in shear flow, obtained from a preliminary simulation.

\section{Numerical results}

\subsection{Identical vesicles in the same shear plane}

\begin{figure}

\includegraphics[width=\columnwidth]{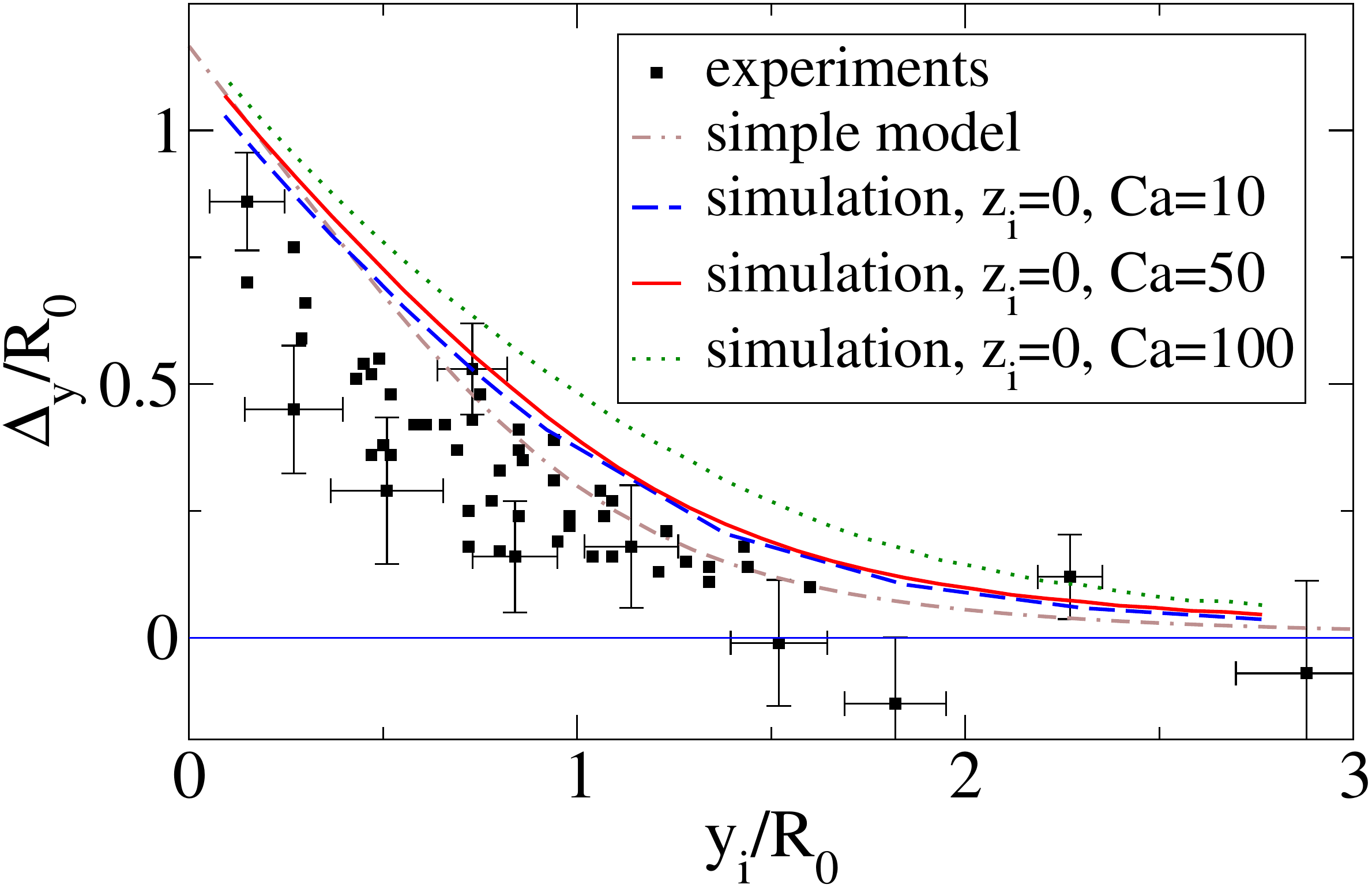}

\caption{{\label{interaction_curve_zi0}Simulated displacements in the $y$ direction for $z_i=0$, $\lambda=1$, $\nu=0.95~(\nu_a=0.91)$, $Ca\in\{10,50,100\}$, and comparison with experimental data (same data as in Fig. \ref{fig:exp-lambda1}).} {Error bars correspond to a spatial uncertainty of $1$\,\textmu m. Only part of them are presented for clarity. For comparison, we also plot the curve obtained through our simple model (Eq. \ref{eq:emplaw}) with $\xi=8U$ given by the $U$ value obtained from Olla's theory \cite{olla97} for $\nu=0.95$.}}

\end{figure}

We start with the case of identical vesicles, with the typical parameters $(\lambda=1,\nu=0.95)$, in the same shear plane of an infinite simple shear flow. The capillary number is taken in the set $\{10,50,100\}$, so that the whole range of possible experimental values is covered. We plot in Fig. \ref{interaction_curve_zi0} the interaction curve $\Delta_y(y_i)$ (that is the difference between the final and initial $y$-positions), with initial and final distances corresponding to $x_{i}=-10a$ and $x_{f}=10a$.

\textcolor{black}{For all values of $Ca$, we recover the  decrease of $\Delta_y/R_0$ from around $1$ to zero. All deviations become smaller than $0.1$ for $y_i/R_0>2.5$.  There is  a  good agreement with the simple model based on the law for the lift of a vesicle near a wall, that was presented in the preceding section. It thus validates this model as a convenient tool to anticipate the dependency of the lift with the mechanical properties of the vesicles. Note however that, since the shape in Olla's model is prescribed, no dependency on $Ca$ can arise from it.}

\textcolor{black}{Overall, considering that there are no fitting parameters (but some experimental uncertainty on $C_a$), the agreement is rather satisfactory. However, the experiments lead to smaller  displacements, as the numerical curve passes through the error bars of \textcolor{black}{only} about $30\%$ of the experimental points. This discrepancy may be explained by differences between the experimental configuration and the ideal unbounded simple shear flow on three aspects. First, the suspension is slightly polydisperse, both in shape and size. Second, the centering in the $z$ direction might not be perfect. Indeed, the  depth of focus of the microscope is about $5$\,\textmu m, which allows $z_i$ to differ from $0$ by amounts up to $R_0/2$. Third, the balance between wall-induced lift forces and sedimentation in the $y$ direction is  perturbed during the interaction, and may also not be fully reached before interaction starts, because of preceding interactions. All those effects could be non negligible.
We use the numerical model to study their relative importance. }

\subsection{Departure from interaction  of two identical vesicles in a shear plane of a simple shear flow}

\subsubsection*{Influence of polydispersity}

Regarding the influence of polydispersity, we computed several sets of interaction curves, with $Ca=10$, first changing the radius ratio so that $R_2/R_1\in\{0.9,1.1\}$, and then both reduced volumes, so that $\nu_1=\nu_2\in\{0.8,0.99\}$. We find relatively small effects, not sufficient to explain all the data scattering : for instance, for $y_i/R_0=0.5$, the maximal variation in $\Delta_y$  is $9\%$. Such a small effect was expected from the qualitative discussion presented in Sec. \ref{experimental_results}.

\subsubsection*{Influence of $\mathbf{z_i}$}

\begin{figure}[h!]

\centering

\includegraphics[width=\linewidth]{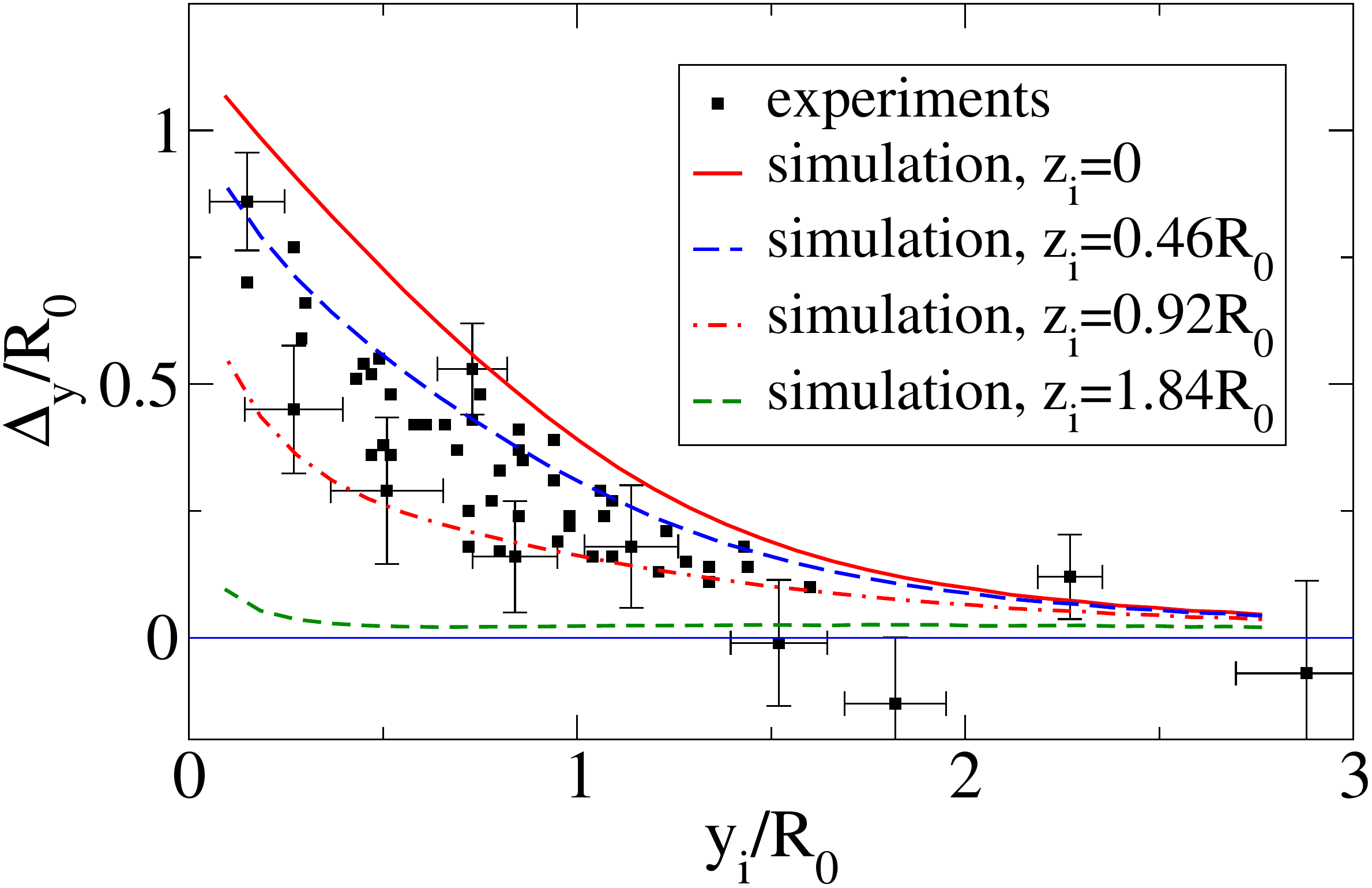}

\caption{\label{interaction_curve}Simulated displacements in the $y$ direction for $z_i\in\{0,0.46R_0,0.92R_0,1.84R_0\}~(0.92R_0=a)$, $\lambda=1$, $\nu=0.95~(\nu_a=0.91)$, $Ca=50$, and comparison with experimental data (same data as in Fig. \ref{fig:exp-lambda1}).}

\end{figure}

We plot the interaction curve $\Delta_y(y_i)$ for $z_i\in\{0,0.46R_0,0.92R_0,1.84R_0\}~(0.92R_0=a)$. The result is reported in Fig. \ref{interaction_curve}.

{As expected, the deviation $\Delta_y$ decreases with $z_i$. Considering that the vesicles can be initially shifted in the vorticity direction by the maximal distance allowed by the focal depth of the microscope, a better agreement between experimental data and simulations is found (about 70\% of experimental points, for vesicles of radii 10 \textmu m).}

\subsubsection*{{Influence of the bottom channel wall}}

We consider now the influence of an imbalance between wall-outward migration and sedimentation. For simplicity and since gravity acts similarly on both vesicles, we only consider  the wall migration effect. As lift is a decreasing function of the distance to the wall \cite{coupier08}, we expect the upper vesicle to migrate less relatively to the wall, so that the distance between the two vesicles is indirectly reduced due to that wall-induced lift forces.

We compare the whole trajectory obtained by our code with the one  corresponding to the experiment shown on Fig. \ref{fig:trajexp}. The geometrical input parameters of the simulation are the reduced volume $\nu$ and the 3D effective radius $a$, in contrast with the experimentally measured reduced area $\nu_a$ and the 2D effective radius $R_0$.  From the study of isolated vesicles in simple shear flow, we find that, for $Ca=10$,  vesicles having same 2D cross-sections as the vesicles of Fig. \ref{fig:trajexp}  are characterized respectively by $\{\nu=0.98, a= 8.9\text{\,\textmu m}\}$  and $\{\nu=0.97, a= 8.6\text{\,\textmu m}\}$.

\begin{figure}[t!]

\begin{center}

\includegraphics[width=\linewidth]{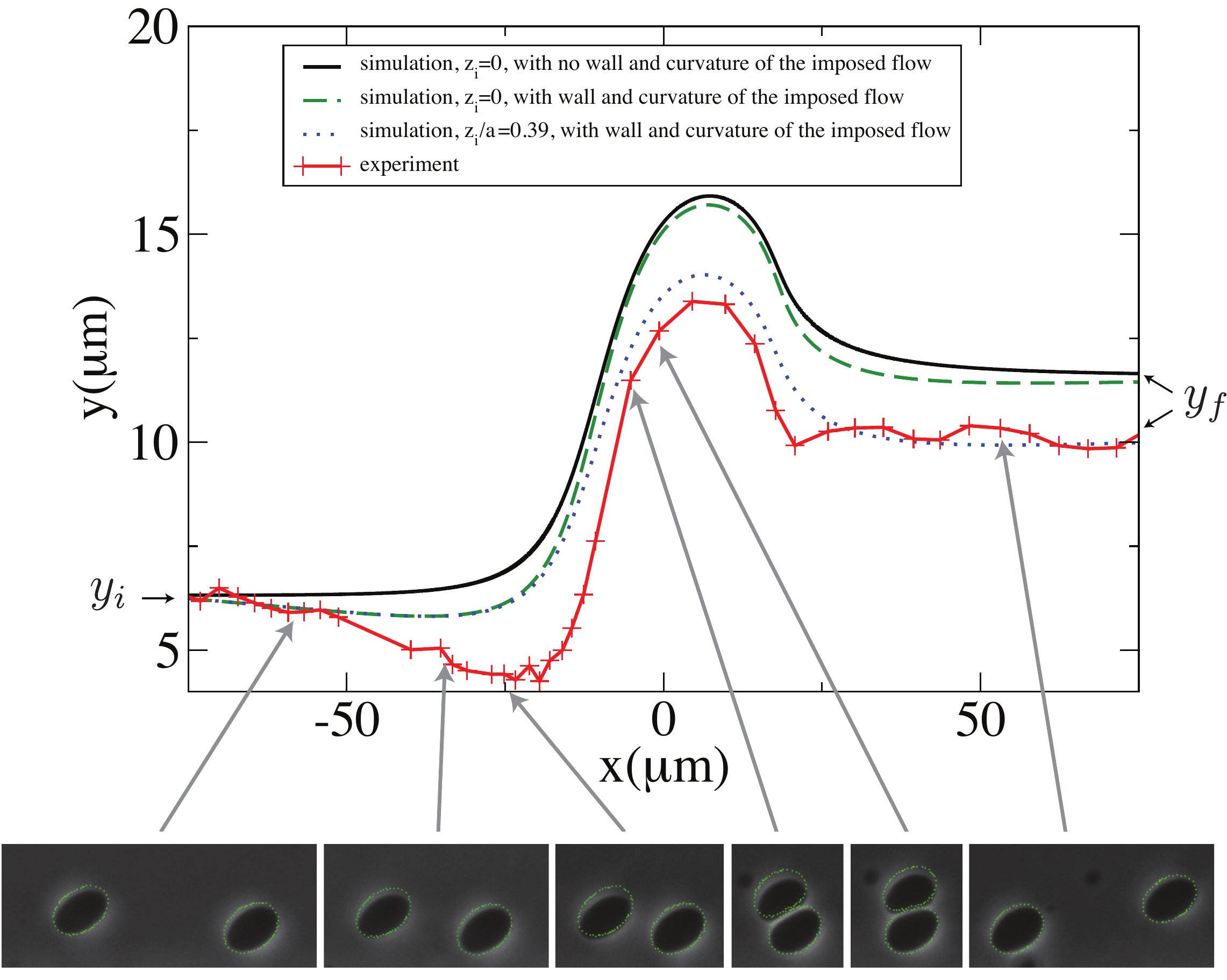}

\caption{\label{influ_curv}Influence of the presence of the wall and the curvature of the imposed flow:  comparison between the trajectory $y(x)$  of the experiment of Fig.  \ref{fig:trajexp} and the simulations, where we consider $Ca=10$. {For six selected relative positions $x$, the raw experimental pictures are shown, on which  the simulated shapes are superimposed for the case where the influences of the wall and curvature of the imposed flow are considered,  and $z_i/a=0.39$ (dotted line).}}

\end{center}

\end{figure}

In order to quantify the bottom wall effect, we adopt the Green's function corresponding to a semi-infinite fluid \cite{Pozrikidis92,kaoui09_1}, and include the quadratic part of the flow in the plane of shear, so that the imposed flow is $\dot{\gamma}y(1-{y}/{L_y})\boldsymbol{e}_{x}$. The initial distance of vesicle 1 from the wall is $y_{i,1}=32$\,\textmu m. The comparison between the experiment and the numerical study  is presented in Fig. \ref{influ_curv}, without and with wall, for vesicles in the same shear plane ($z_i=0$). A possible shift $z_i/a=0.39$ is also considered together with the presence of the wall.

As expected, lift by the wall leads to a slight initial attraction (a decrease of $y$ for $x<0$), which results in a slightly smaller final lateral displacement when $x\to\infty$.
It appears however that this correction is too small to account for the remaining discrepancy between the simulations and the experiments, for which the initial attraction of around 1\,\textmu m, that is seen on Fig. \ref{fig:trajexp}, appears on most trajectories. 

Anyhow, this second-order  effect  is most probably linked with the presence of the wall, as suggested by recent simulations by Narsimhan et al., where interacting red blood cells in the vicinity of a wall are studied \cite{narsimhan13}. They show that, for particles close enough to a wall, the relative lateral distance might decrease before interaction (see trajectories on their Fig. 15(b)), sometimes even leading to a completely different scenario of interaction involving swapping trajectories where particles do not cross. \textcolor{black}{As shown by the authors, the presence of the bottom wall in the y direction induces the formation of a recirculation vortex behind the first  particle, which is mainly responsible for the initial attraction. It is likely that when walls are also present in the z direction, as is the case in the experiment, the strength and extension of this recirculation are larger, leading to the stronger attraction observed in the first  stage of experimental trajectories.}

To sum it up, starting from comparable results for experiments and simulations, we have shown that a shift in the vorticity direction and the contribution of walls, both being inherent to the experiment, lead to a decrease of the repulsion, thus to some scattering in the experimental data, that all lie right below the ideal curve given by the simulations.

\subsection{Deflection in the vorticity direction}

As an extension to experimental results, the model also allows to investigate the effect of the interaction on the deflection $\Delta_z$.

\begin{figure}[h!]

\centering

\includegraphics[width=\linewidth]{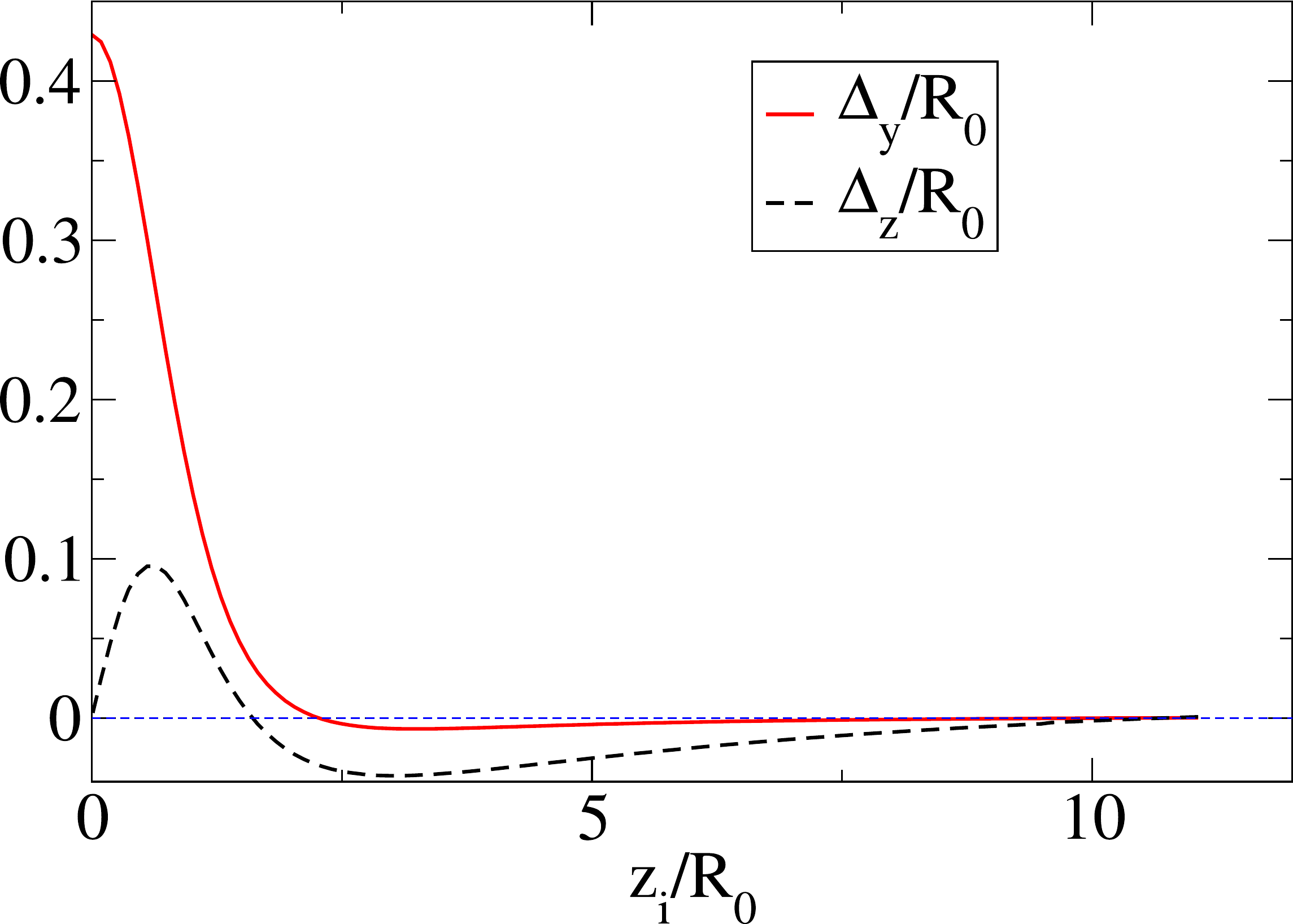}

\caption{\label{interaction_curve_z}Interaction curves for  $\Delta_y({z_i}/{R_0})$ and $\Delta_z({z_i}/{R_0})$, for $y_i=0.92R_0=a,\lambda=1,\nu=0.95,Ca=50$, and $|x_i|=|x_f|=20a$}

\end{figure}

In Fig. \ref{interaction_curve_z}, we present the interaction curves $\Delta_y({z_i})$ and $\Delta_z({z_i})$, for $y_i=0.92 R_0$, the other parameters remaining the same as previously.

We find that there is a range of initial transverse positions for which the interaction leads to a transverse attraction between the vesicles, mostly in the vorticity direction. A similar phenomenon has been predicted for the interaction of capsules \cite{lac08}, but not for drops \cite{loewenberg97}. An asymptotic study, for vesicles in the far field interacting regime, also predicts such an attraction \cite{gires12}. However, here the vesicles become close during the interaction, so a qualitative interpretation of the predicted attraction may involve the description of the flow of the thin liquid \textcolor{black}{film} between the two tank-treading membranes, as done for drops \cite{loewenberg97}.

\section{Hydrodynamic diffusion}

From the numerical study, one can expect to deduce results about the hydrodynamic diffusion properties of vesicle suspensions, in a regime  where the solution is concentrated enough so that interaction effects are not negligible, but dilute enough so that pairwise interactions dominate over three-body interactions. We mostly study the case of self-diffusion, a phenomenon related to the average transverse motion of a single vesicle. We also find that an estimation of the collective diffusion coefficients is not possible only considering two-vesicle interactions, due to the long range of hydrodynamic interactions.

\subsection{Self-diffusion}

\subsubsection{Theoretical background}

We consider a homogeneous suspension of vesicles, described at a mesoscopic level by a volume fraction $\phi$. For a given initial state, if this suspension is sheared by an imposed flow $\mathbf{v}^{\infty}=\dot{\gamma}y\mathbf{e}_x$, a given vesicle will interact with the others and, as a result, will undergo a net displacement $\mathbf{X}$ from its original streamline. In an unstructured semi-dilute suspension, the transverse motion of the vesicle is expected to be a random walk due to successive interactions with different vesicles. At long times, its mean-squared displacement $\langle X_\alpha^2 \rangle$ is described by the self-diffusion coefficients $D_{s,\alpha}$, defined by

\begin{eqnarray*}
D_{s,\alpha}&=&\lim_{t\rightarrow\infty}\frac{1}{2}\frac{d\langle X_\alpha^2\rangle}{dt},
\end{eqnarray*}
with $\alpha\in\{y,z\}$, $\langle.\rangle$ being an ensemble average  over all possible initial states of the suspension\cite{loewenberg97}.

As detailed by Da Cunha and Hinch \cite{dacunha96}, assuming that only two-vesicle interactions occur leads to the following expression for $D_{a,\alpha}$:

\begin{equation}
\label{expr_self_diff}
 D_{s,\alpha}=\phi \dot{\gamma}a^2f_\alpha,
\end{equation}

with

\begin{equation}
\label{def_f}
f_{\alpha}={\frac{3}{2\pi}}\int_{(y_i,z_i)\in[0,+\infty]^2}\Delta_{l,\alpha}^2y_idy_idz_i,
\end{equation}

where $\Delta_{l,\alpha}$ is the deviation of a test vesicle  in the laboratory frame after one interaction.  For identical vesicles, $\Delta_{l,\alpha}=\Delta_{\alpha}/2$. In the latter integral and from now on, all lengths are expressed in $a$ units.

\subsubsection{Analysis of the formal convergence of the diffusion coefficient}
\label{formal_convergence}

As the hydrodynamic interaction between two vesicles slowly decreases, the question of the convergence of the previous integral arises. As all displacements $\Delta_{l,\alpha}$ are bounded, the convergence of the expression \ref{def_f} is linked to the contribution of the integration domain $\sqrt{y_i^2+z_i^2}\gg 1$. We analyze this contribution by using an asymptotic study  of two interacting quasi-spherical vesicles remaining very distant from each other, that was recently proposed by Gires \textit{et al.} \cite{gires12}.  We first need to determine the domain in the $(y_i,z_i)$ space for which this asymptotic study is valid, a discussion that was not provided in the original paper. For the asymptotic study to be valid, vesicles must remain far enough along the whole trajectory, so that, at all times, $||\mathbf{R}_{12}||\gg1$. As $\sqrt{y_i^2+z_i^2}\gg 1$, one could expect this criterion to be always satisfied. However, let us consider $y_i=0$. If there was no interaction, both vesicles would flow with the same velocity. But, since the velocity field induced by one vesicle is radial, vesicle collision may occur, which is inconsistent with the asymptotic approach. These considerations hint to the fact   that the asymptotic study may not be valid for $y_i\ll1$.

In order to get a more accurate validity criterion, we assume the asymptotic study to be valid for all times, and check that the  inter-vesicle distance remains large. We expect that this approach can be used as each vesicle is not in the vicinity of a bifurcation phenomenon, such as the transition between the tank-treading and vacillating-breathing modes.

As detailed in Gires \textit{et al.}\cite{gires12}, within this asymptotic approach with respect to the inter-vesicle distance, the trajectory of $C_2$ with respect to  $C_1$ is of the form:

\begin{eqnarray}
\label{y(x)}
y(x)= & y_i\nonumber +\frac{1}{\dot{\gamma}}\left[\left(\frac{x^3}{(x^2+b^2)^{3/2}}+1\right)\frac{T_{xx}}{b^2}-\frac{2y_iT_{xy}}{(x^2+b^2)^{3/2}}\right.\nonumber\\
& \left.+\left(\frac{(2x^2+3b^2)x}{(x^2+b^2)^{3/2}}+2\right)\frac{y_{i}^2T_{yy}}{b^{4}}\right],
\end{eqnarray}

and

\begin{equation}
\label{z(x)}
z(x)=\frac{(y(x)-y_i)z_i}{y_i}+z_i,
\end{equation}
where $b=\sqrt{y_i^2+z_i^2}$, and $\{T_{xx},T_{xy},T_{yy}\}$ are constants linked to the perturbation of the velocity field induced by the vesicles.

As the symmetry of the system does not depend on the reduced volume, we expect these scalings to be valid for vesicles of arbitrary deflation in the tank-treading regime, the dependency on the reduced volume being accounted for  by the tensor $T_{\alpha\beta}$.

As $y(x)-y_i=O(b^{-2})$, the $y$ distance between the vesicles will remain large if initially large. However, as $z(x)-z_i={z_i}/{y_i}\times O(b^{-2})$, problems may arise at small $y_i$, as discussed earlier. In this case, the prevalent term in Eq. \ref{y(x)} is the term proportional to ${T_{xx}}/{b^2}$. If $T_{xx}<0$, this could lead to a minimal distance in the vorticity direction of the form  $z_{\text{min}}=z_i-{c}/{(y_iz_i)}$, with $c>0$. In order that the asymptotic approach remains valid starting with $z_i\gg1$, we impose the condition that $z_{\text{min}}>{z_i}/{d},\quad(d>1)$, where $d$ is a constant. This criterion can be achieved if $y_i>{e}/{z_i^2}$, with $e=dc/(d-1)>0$.

For initial positions satisfying this criterion, we find from Eqs. \ref{y(x)} and \ref{z(x)} that

\begin{eqnarray}
\label{inter_dy}
\Delta_y=O(\frac{y_i^2-z_i^2}{(y_i^2+z_i^2)^2}),\\
\label{inter_dz}
\Delta_z=O(\Delta_y\frac{z_i}{y_i}).
\end{eqnarray}

It is clear from these expressions that the integral of Eq. \ref{def_f},  restricted to the region where the asymptotic expression is valid, is convergent.

As for the region of large $z_i$ and small $y_i$ with $y_i<{e}/{z_i^2}$, where the asymptotic expression is not valid,  since $\Delta_{l,\alpha}^2$ is bounded by its maximal value and the integral of $y_i$ on this region is finite,  its contribution  to the integral in Eq. \ref{def_f} is bounded, and finally the whole integral is convergent.

\begin{figure}[h!]

\centering

        \includegraphics[width=0.65\columnwidth]{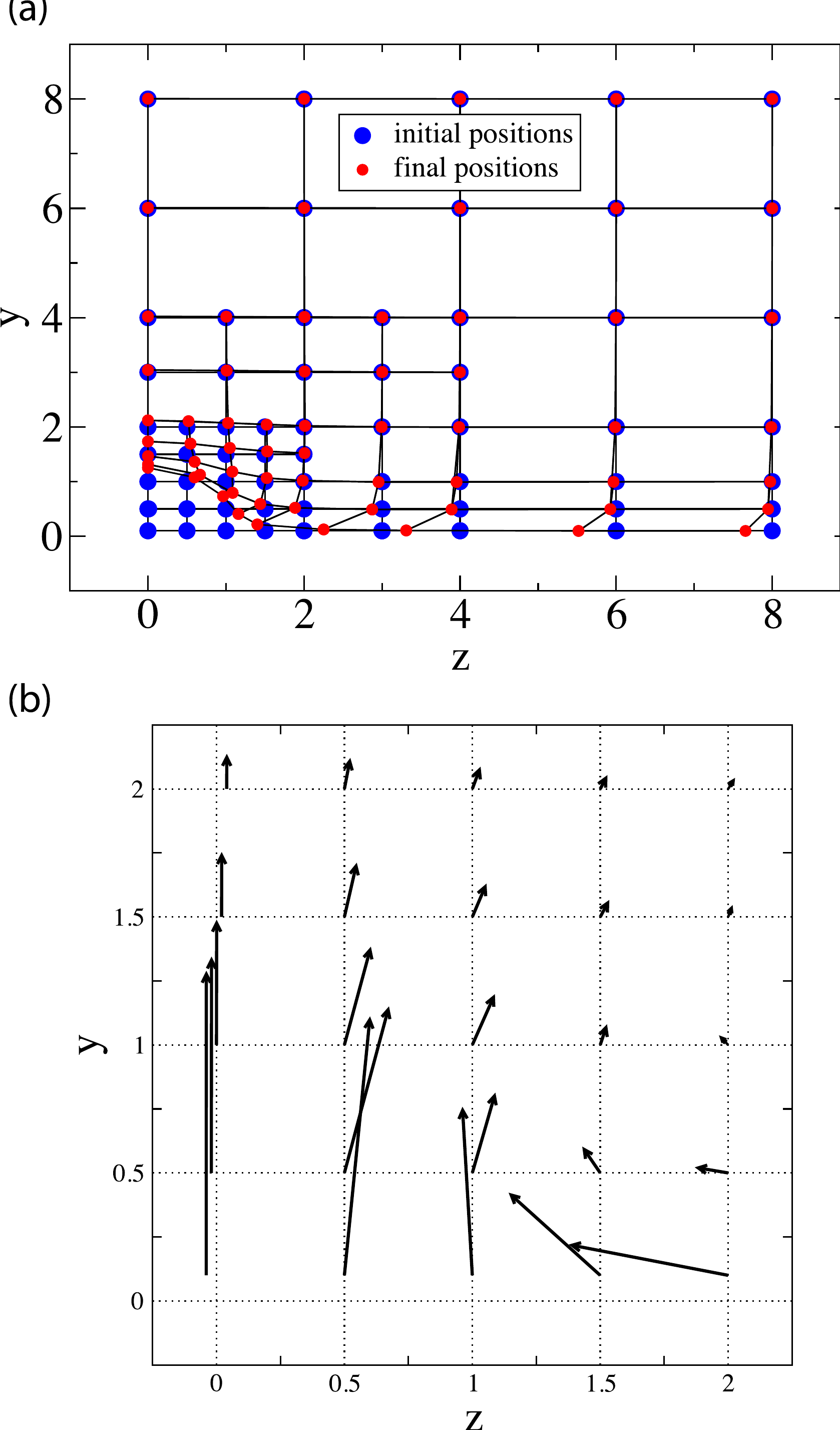}

 \caption{\label{evol_pos_tran} (a) Evolution of the transverse position of a vesicle as a function of $(y_i,z_i)$. $\nu=0.95, Ca= 50$. The lowest considered value for $y_i$ is $y_i=0.1a$, for which we chose $|x_i|=|x_f|=80a$. For the other points, $|x_i|=|x_f|=40a$. Panel (b) shows a zoom on the range $(y_i,z_i)\in[0,2]\times[0,2]$, where displacements are represented by arrows. For $z_i=0$, arrows are slightly shifted to avoid superimposition.}

 \end{figure}

\subsubsection{Numerical determination of the self-diffusion coefficient}
We now estimate the value of $f_y$   (Eq. \ref{def_f}) that enters the expression of the diffusion coefficient (Eq. \ref{expr_self_diff}). For that purpose we need to run several simulations by starting with different initial position in the $y-z$  plane (which is the plane orthogonal to the flow direction) and determine by how much the initial relative positions $y_i$ and $z_i$ have varied (by amounts $\Delta_y$ and $\Delta_z$) after the two vesicles have interacted. We discretize the domain of initial values $(y_i,z_i$)  by considering the following domain size 
 $[0,8]\times[0,8]$ (in units of vesicle radius $a$). The discretized lattice of initial positions is shown in Fig. \ref{evol_pos_tran}(a) with dark gray disks (blue online). Since the interaction is important only when the two vesicles are separated by about 2 or 3 radii, the lattice has a wide enough periodicity far away from 
 $(0,0)$, whereas in the vicinity further refinements are chosen in order to gain numerical precision. More precisely, the domain is decomposed into three regions  $A$, $B$ and $C$, consisting in $[0,2]\times[0,2]$, $\{[0,4]\times[0,4]\}\diagdown\{[0,2]\times[0,2]\}$ and $\{[0,8]\times[0,8]\}\diagdown\{[0,4]\times[0,4]\}$. 
The lighter gray disks (red online) in Fig. \ref{evol_pos_tran}(a) show the final relative positions $y_f$ and $z_f$. We note that in region C the effect is weak, while it becomes more and more pronounced in region $B$ and $A$.
The contributions of the integral involved in Eq. \ref{def_f} on the different sub-domains $A$, $B$ and $C$  are then evaluated using a trapezoidal rule. The results are given in Table \ref{table1}.

    \begin{table}[t!]

\begin{center}

 \begin{ruledtabular}

 \begin{tabular}{ccc}

   & $f_y$ & $f_z$  \\

\hline

  part A & 0.028 &  0.002  \\

\hline

part B & 0.003 &  0.004  \\

\hline

part C & 0.001 &  0.005  \\

\end{tabular}

  \end{ruledtabular}

  \caption{\label{table1}Contributions of the sub-domains to the dimensionless self-diffusion coefficients $f_y$ and $f_z$. $\nu=0.95, Ca=50$.}

\end{center}

\end{table}


We find that the contributions for $f_y$ are decreasing. As we proved the convergence of the expression, we expect the contributions of the remaining part of the plane to be at most of the order of the contribution of the sub-domain C, and thus get the following estimation for $f_y$:

\begin{equation}
 f_y=0.032\pm10\%.
\end{equation}

The uncertainty of $10\%$ is a rough estimate coming from a study of the sensitivity of the code to some numerical parameters, like a tension parameter used to preserve locally the area of the membrane.

For $f_z$, we do not get decreasing contributions, due to the slow decrease of $\Delta_z$ with $z_i$ when ${y_i}\ll1$. {A similar study has been presented by Zhao and Shaqfeh \cite{zhao13}}, {who calculated $f_y$ for $\nu=0.95$ and $Ca=1$. Using the effective radius  based on the surface as a length scale ($a'=\sqrt{S/4\pi}$, where $S$ is the vesicle membrane area) they find $f_y=0.028$. With the same convention instead of our choice of radius based on the volume, we find $f_y=0.032 \nu^{2/3}=0.031$, for $\nu=0.95$ and $Ca=50$, which is a consistent result since lateral displacement increases with $Ca$ (Fig. \ref{fig:exp-lambda-autre}). Zhao and Shaqfeh also estimated the value of $f_z$, restricting to the integration domain $[0,3]\times[0,3]$ : their value matches ours on the same region. However, the present study shows that restricting the integration to this domain is not sufficient to get a quantitative value of $f_z$, due to the slow decrease of the attraction with $z_i$ for vesicles characterized by $y_i\ll1$.}

We are not aware of experimental measures of $f_y$ to which we could  compare our estimation. On the basis of studies on suspensions of spheres, the assumption of considering only two-vesicle interactions could be a good approximation up to volume fractions of about $10\%$ \cite{Larson99}.

\subsubsection{Discussion}

From simulated trajectories, Loewenberg and Hinch \cite{loewenberg97} calculated $f_y$ and $f_z$ for pairs of drops as a function of viscosity ratio and capillary number. They evoke the scaling at long distance $\Delta_{\alpha} \sim 1/(y_i^2+z_i^2)$, which is similar to ours, to prove the convergence of the integral of Eq. \ref{def_f}. It appears that in the case of drops, restricting the integration domain to A+B is sufficient, for $f_y$ as for $f_z$. For $\lambda=1$, $f_y$ was found to be around $0.03\pm 0.01$, depending on the capillary number, a result close to ours. A more quantitative comparison is precluded by the dependency with capillary number and the difference in nature between the elastic restoring forces involved in drops and vesicles. Interestingly, Loewenberg and Hinch \cite{loewenberg97} find $f_z \simeq 0.004$, while we already find a result 3 times larger by integration over A+B+C. We can conclude that anisotropy in self-diffusion is lower for vesicles than for drops. This weaker anisotropy is also stated by Lac and Barth\`es-Biesel in their study of capsules collisions, though $f_y$ and $f_z$ are not calculated \cite{lac08}.

\subsection{Down-gradient diffusion}

The collective diffusion property of a vesicle suspension can be modeled in the following way: we consider a suspension of vesicles with a linearly increasing concentration given by $\phi=\phi_0+\alpha y$, sheared by an imposed flow $\mathbf{v}^{\infty}=\dot{\gamma}y\mathbf{e}_y$. As a result of the hydrodynamic interactions between the vesicles, we expect a collective diffusion of the vesicles to appear, consisting of a transverse flux $\mathbf{j}=j\mathbf{e}_{y}$ of vesicles. As for molecular diffusion due to thermal motion, $\mathbf{j}$ is expected to be in the opposite direction of $\mathbf{\nabla}\phi$, of order $O(\alpha)$. Thus, for $\frac{\alpha a}{\phi_0}\ll1$, we expect that $j=-D_{c,y}\alpha$, with $D_{c,y}>0$. As done by Da Cunha and Hinch  in the case of rough spheres \cite{dacunha96}, we tried to estimate $D_{c,y}$ assuming only two-vesicle interactions. This leads to an expression involving the integral of $y_i^2\Delta_y$ over the plane. However, as $\Delta_y=
 O(\frac{y_i^2-z_i^2}{(y_i^2+z_i^2)^2})$ and the integral of $\frac{y_i^2(y_i^2-z_i^2)}{(y_i^2+z_i^2)^2}$ over $[0,y_0]\times[z_0,+\infty]$, with $(y_0,z_0)\in {\mathbb{R}^{*}_+}^2$, is divergent, it turns out that the estimated expression is not convergent.  A renormalization procedure, analogous to the one used by Batchelor \cite{Batchelor72b,batchelor72c}, and followed  by Wang \textit{et al.}\cite{Wang98}  in the case of the study of the hydrodynamic diffusion properties of a suspension of spheres, may lead to a convergent expression. It is hoped to investigate this matter further in a future work.

\section{Conclusion}

We performed an experimental and numerical study of the trajectory deviations of identical vesicles interacting in shear flow. In experiments, restricted to pairs of vesicles in the same shear plane, the amplitude of the net displacement decreases quickly when the initial \textcolor{black}{lateral} distance increases and becomes negligible when this  distance is larger than approximately two vesicle radii. 

A simplified model based on the well established law for the lift of a vesicle near a wall was proposed, which allows to estimate quantitatively how the displacement should vary with the mechanical properties of the vesicles.

\textcolor{black}{With no fitting parameter, the deviations are found to be in rather good agreement with our 3D simulations,}
\textcolor{black}{ even if smaller deviations are found experimentally. We found than the main part of this discrepancy can be due to }
\textcolor{black}{differences between the experimental configuration and the ideal case of unbounded shear flow where the two vesicles would be perfectly coplanar. The effect of walls, recently highlighted by Narsimhan et al. \cite{narsimhan13}, would need to be quantified thoroughly in complementary experiments where our requirements of similar deflation within the pair of vesicles could be loosened for  simplicity, since the effect of deflation has been characterized and shown to be weak. We also indicate that, according to partial results not shown here, the requirement of identical size within a pair may be released, as rescaling of the displacements by the average radius $R_0$ of  two vesicles  of different size lead to \textcolor{black}{a similar} curve for $\Delta_y/ {R_0}$ as a function of $y_i/{R_0}$.}

In addition, displacements in the vorticity direction were explored through the simulations and found to be about an order of magnitude lower than in the shear direction, with a range of initial distances leading to a weak attraction of vesicles.

Shear-induced diffusion coefficients can be obtained by a proper averaging of the net displacement over all initial configurations. The self-diffusion, related to the random walk of vesicles in a suspension, can be quantified using a discrete integration over a relatively small domain for the diffusivity in the shear direction, and could be determined in the vorticity direction if a larger integration area was considered, due to the slower decrease of the amplitude of displacement in that direction. Note that this integration would not have been possible in 2D \cite{li00} where the displacements would scale like $1/y$ instead of $1/y^2$.

An estimation of the down-gradient diffusivities as defined by Da Cunha and Hinch \cite{dacunha96} was not possible due to the long range of hydrodynamic interactions, leading to a divergence of the integrals. In this case, the dilute limit assumption breaks down and one can no longer consider only pair interactions as is the case for rough spheres with short range interactions \cite{dacunha96}.

\appendix

\section{Local conservation of the area}

\label{appendixA}

\begin{figure}[h!]

\begin{center}

\includegraphics[width=\linewidth]{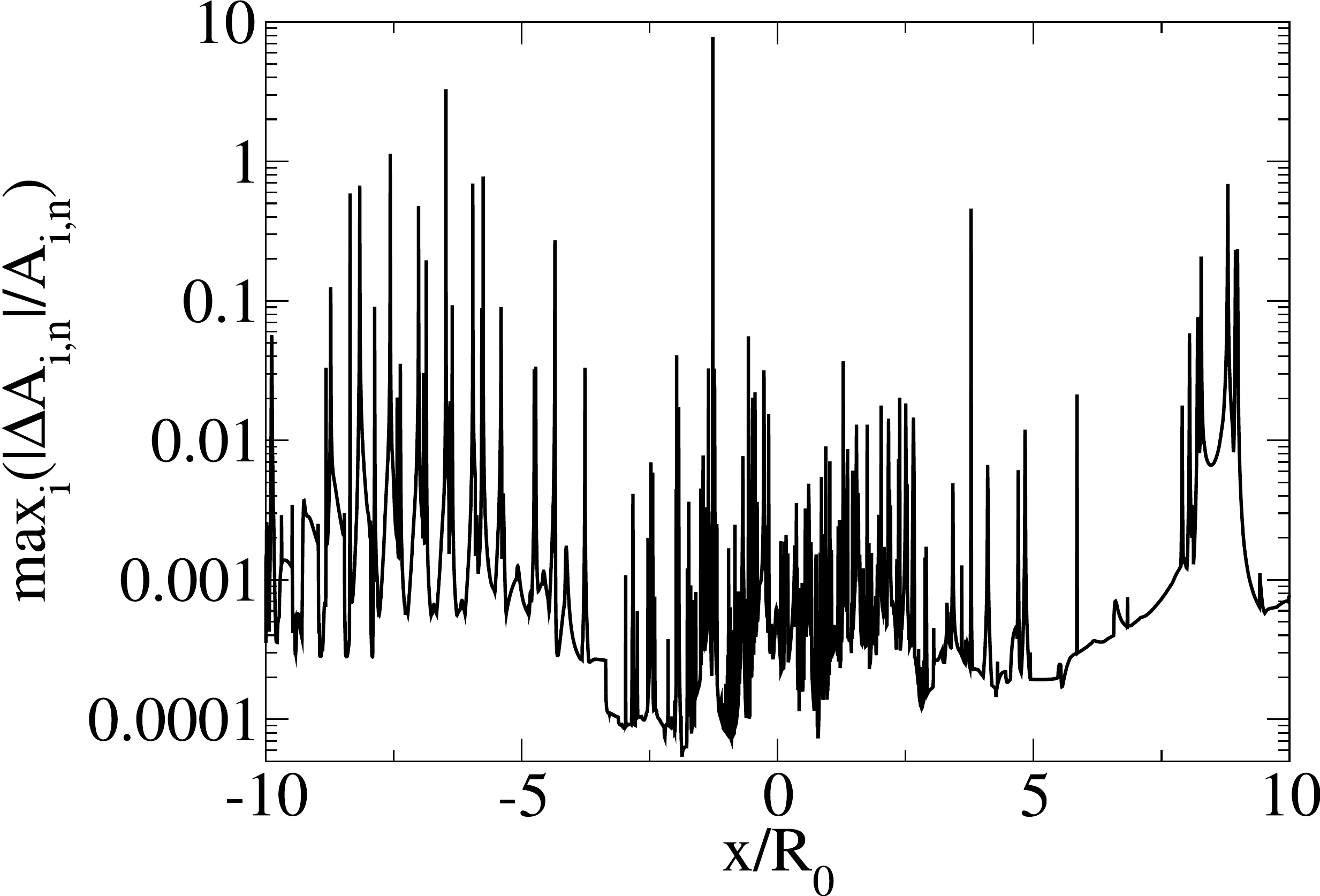}

\caption{\label{trace_max}Maximal relative variation of the area of the faces. $A_{\text{i},n}$ is the area of face $\text{i}$ at time step $n$, and $\Delta A_{\text{i},n}$ is its variation during one time step, if it were advected by the full velocity field.}

\end{center}

\end{figure}

We present in this appendix results about the local conservation of the area of a vesicle, during a typical trajectory. The parameters chosen were $(Ca=50,\lambda=1,\nu=0.95,y_i=0.5 a,z_i=0)$. First, we plot in Fig. \ref{trace_max} the maximal relative variation of the area of the mesh faces, between two time steps, if they were advected by the full velocity field (in the simulation, the vertices are only advected by the normal component of the velocity field).

\begin{figure}[h!]

\begin{center}

\includegraphics[width=\linewidth]{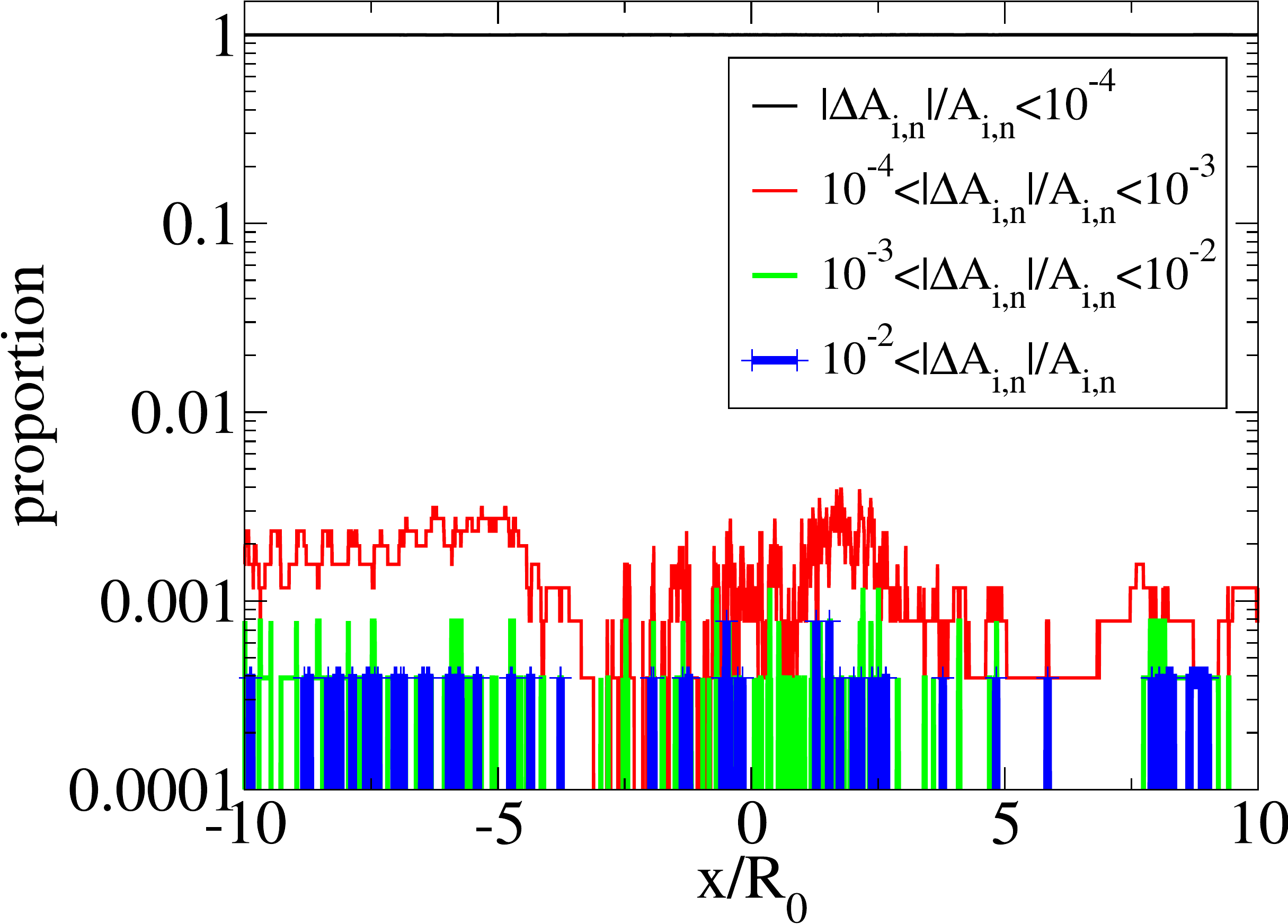}

\caption{\label{proportion}Proportions of faces whose relative variation are in the ranges $[0,10^{-4}],~[10^{-4},10^{-3}],~[10^{-3},10^{-2}]$ and $[10^{-2},+\infty]$.}

\end{center}

\end{figure}

We find that this maximum can reach values higher than one. However, the proportion of faces corresponding to such values stays lower than $0.1\%$, as shown in Fig. \ref{proportion}.

\begin{acknowledgments}

The authors thank T. Biben (LPMCN, Lyon, France) for providing the initial simulation code and CNES (Centre National d'Etudes Spatiales) and ESA (European Space Agency) for financial support. P.-Y.G. also acknowledges funding from the French Ministry of Higher Education and Research, and the CNRS (Centre National de la Recherche Scientifique).

\end{acknowledgments}

\end{document}